\documentclass{phbauth}
\usepackage{graphicx}

\begin{document}

\begin{frontmatter}

\title{
Critical temperature and phase diagram for a 2D superfluid Fermi-gas with
repulsion.
}

\author[address1]{D.V. Efremov},
\author[address1]{M.S. Mar'enko\thanksref{thank1}},
\author[address2]{M.A. Baranov},
\author[address1]{M.Yu. Kagan}

\address[address1]{P.L.Kapitza Institute for Physical Problems, Moscow 117334,
Russia}
\address[address2]{
    Russian Research Center "Kurchatov Institute", Moscow
123182, Russia
}

\thanks[thank1]{Corresponding author. Present address:
    P.L.Kapitza Institute for Physical Problems, Kosygin str. 2, Moscow
117334, Russia.
E-mail: maxim@kapitza.ras.ru}

\begin{abstract}
In the framework of a 2D Fermi-gas with short-range repulsive interaction we
find the critical temperature of the superfluid phase transition based on
Kohn-Luttinger effect, and analyze its dependence on
magnetic field.  We calculate strong-coupling corrections to the
Ginzburg-Landau free energy functional and
analyze the staibility of 2D superfluid phases. Possible
applications of the results to real systems are discussed.
\end{abstract}


\end{frontmatter}

\section{Introduction}
Recent experiments on $^3$He submonolayers absorbed on graphite
\cite{Greywall}, $^3$He atoms bound to the surface of
$^4$He films \cite{Saunders,Hallock,Godfrin} and discovery of
superconductivity in layered HTSC materials stimulate theoretical studies of
superfluid phase transition in a 2D Fermi liquid.
It was shown \cite{KohnLutt,KaganChub}, that even in
the case of a bare repulsive $s$-wave interaction in a 3D Fermi system, there is
a transition to the $p$-wave superfluid state due to a singularity in the
effective interaction near the momentum transfer $q=2p_\mathrm{F}$
(Kohn-Luttinger effect), $\Gamma_\mathrm{sing}(q) \sim (2
p_\mathrm{F}-q) \ln |2 p_\mathrm{F}-q|$, where $p_\mathrm{F}$ is the Fermi
momentum.  Further analysis \cite{BCK} shows that the similar effect takes
place also in 2D case.

\section{Theoretical model and results}
We consider a 2D non-ideal Fermi gas with a short-range repulsive
interaction, $r_0p_\mathrm{F} \ll 1$, where $r_0$ is the range of
potential, such that the perturbative expansion in a gas parameter $f_0$,
related to the $s$-wave scattering amplitude  between two particles on the
Fermi surface, is legitimate.  For the considered case the gas parameter is
given by
  \[
  f_0=\left[4\pi/mU_0 + \ln (r_0 p_\mathrm{F})^{-2}\right]^{-1},
\]
where $U_0$ is the $s$-component
of the potential.

It turns out that for a 2D Fermi gas the Kohn-Luttinger mechanism works in the
third order in $f_0$, because the singular contribution of
the second order in $f_0$ to the effective interaction,
\[
({m}/{4\pi})\Gamma_\mathrm{sing}^{(2)}(q) = - f_0^2 \mathrm{Re}
\sqrt{1- (2p_\mathrm{F}/q)^2}, \]
is zero for $q \le
2p_\mathrm{F}$, and, therefore, does not contribute to the $p$-harmonic. The
        result for $\Gamma_\mathrm{sing}(q)$ in the third order in $f_0$ is
  \[
({m}/{4\pi})\Gamma_\mathrm{sing}^{(3)}(q) \sim - f_0^3
\mathrm{Re}\sqrt{1-(q/2 p_\mathrm{F})^2}, \] 
and it leads to an
  effective attraction in the $p$-wave channel. Numerical calculations of
third-order diagrams give for the
$p$-wave critical
temperature
\begin{equation} T_\mathrm{c1} =
    \tilde{\varepsilon} \exp \left(-\frac{1}{6.1 f_0^3} \right),
    \end{equation}
  where
$\tilde{\varepsilon} =  A\cdot \varepsilon_\mathrm{F}$, $A$ is the unknown
numerical prefactor,  $\varepsilon_\mathrm{F}=p_\mathrm{F}^2/2m$ is the
Fermi energy.

The typical value of scattering amplitude
obtained from the experiments on $^3$He atoms on the surface of $^4$He is $f_0
\le 0.3$ for $^3$He coverages corresponding to the dilute Fermi gas
 situation.
In this regime the critical temperature is $T_\mathrm{c1} \sim 10^{-3}K$.

The critical temperature $T_\mathrm{c1}$ can be significantly increased by
applying an external magnetic field, because in this case the effective
interaction contributes to $T_\mathrm{c1}$ already in the second order.
The separation of polarization and pairing effects
leads to a strongly
non-monotonic dependence of the critical temperature on the polarization
(Fig. 1).


\begin{figure}[btp]
\begin{center}\leavevmode
\includegraphics[width=0.7\linewidth]{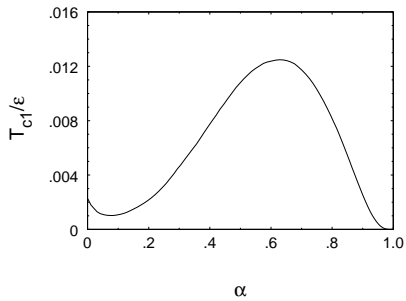}
\caption{
  The $p$-wave critivcal temperature $T_\mathrm{c1}/\tilde{\varepsilon}$ as a
  function of polarization
  $\alpha=\frac{(n_\uparrow-n_\downarrow)}{(n_\uparrow+n_\downarrow)}$ for
  typical $f_0=0.3$.} 
\end{center}\end{figure}

In the weak coupling (BCS) limit for zero magnetic field, two different
phases correspond to the minimum of the free energy (the axial
and the planar phases) \cite{Brusov}.  To lift this degeneracy one has to
consider next order ($\sim T_\mathrm{c}/\varepsilon_\mathrm{F}$) corrections.
Following \cite{RS}, we have calculated the difference between free energies of
axial and planar phases to this order:
   \begin{eqnarray}
     \Phi_\mathrm{axial}-\Phi_\mathrm{planar}
   &\sim& - \frac{N(0)}{T_\mathrm{c}^2} \left(
     \frac{T_\mathrm{c}}{\varepsilon_\mathrm{F}} \right) \\
     && \times \int
    \frac{\mathrm{d}\varphi}{2\pi} |\sin \varphi|
    T_a^2 (\hat{\bf k}_1,\hat{\bf k}_2;\hat{\bf k}_1,\hat{\bf k}_2) ,
    \nonumber
    \end{eqnarray}
where $N(0)$ is the density of states on the Fermi surface,
    $T_a (\hat{\bf k}_1,\hat{\bf k}_2;\hat{\bf k}_3,\hat{\bf k}_4)$ the
    spin-antisymmetric part of quasiparticle scattering amplitude with momenta
    ${\bf k}_i$ on the Fermi surface and $\varphi$ the angle between
    $\hat{\bf k}_1$ and $\hat{\bf k}_2$.
    Clearly, the energy difference (2) is
    always negative, therefore
     in the
    absence of a magnetic field the superfluid Fermi-gas forms the axial phase.

\begin{ack}
The authors are grateful to A.V. Chubukov, G. Frossati, H. Godfrin, R.B.
Hallock, J. Saunders, J. Nyeki and B.P. Co\-wan for stimulating
discussions.  This work has been supported by RFBR grants 98-02-17077,
97-02-16532 and INTAS grant 98-963.  \end{ack}


\begin{thebibliography}{9}
\bibitem{Greywall} D.S. Greywall, Phys. Rev B {\bf 41} (1990) 1842.
\bibitem{Saunders} M. Dann, J. Nyeki, B.P. Cowan, J. Saunders, Phys. Rev Lett.
{\bf 82} (1999) 4030.
\bibitem{Hallock} D.T. Sprague, N. Alikasem, P.A. Sheldon, R.B. Hallock, Phys.
Rev Lett.  {\bf 72} (1994) 384.
\bibitem{Godfrin}  K.-D. Morhard, C.
B\"{a}uerle, J. Bossy, Yu.M.  Bunkov, S.N. Fisher, H. Godfrin, Phys. Rev. B
{\bf 53} (1996) 2658.
\bibitem{KohnLutt} W.  Kohn, J.H.  Luttinger,
Phys.  Rev Lett.  {\bf 15} (1965) 524.
\bibitem{KaganChub} M.Yu.  Kagan,
A.V.  Chubukov, JETP Lett.  {\bf 47} (1988) 525.
\bibitem{BCK} M.A.
Baranov, A.V.  Chubukov, M.Yu. Kagan, Int.  J.  Mod.  Phys.  B {\bf 6} (1992)
2471, A.V.  Chubukov, Phys.  Rev B {\bf 48} (1993) 1097.
\bibitem{Brusov} P.N.  Brusov, V.N.  Popov, JETP {\bf 80} (1981) 1564.
\bibitem{RS} D. Rainer, J.W.  Serene, Phys.  Rev. B {\bf 13} (1976) 4745.
\end{thebibliography}
\end{document}